\documentstyle[preprint,prl,aps]{revtex}

\begin{document}

\title{The surface chemistry of metal-oxygen interactions: \\
a first-principles study of O:Rh(110)}

\author{Kurt Stokbro$^{1}$\cite{KS} and Stefano Baroni$^{1,2}$\cite{SB}}
\address{$^1$Istituto Nazionale di Fisica della Materia (INFM) and \\
Scuola Internazionale Superiore di Studi Avanzati (SISSA), via
Beirut 2/4, I-34014 Trieste, Italy. \\ $^2$Centre Europ\'een de Calcul
Atomique et Mol\'eculaire (CECAM), ENS-Lyon, 46 All\'ee d'Italie,
F-69007 Lyon, France.}

\maketitle

\begin{abstract} We report on a computational study of the clean and
oxygen-covered Rh(110) surface, based on density-functional theory
within the local-density approximation. We have used plane-wave basis
sets and Vanderbilt ultra-soft pseudopotentials. For the clean
surface, we present results for the equilibrium structure, surface
energy, and surface stress of the unreconstructed and $(1\times 2)$
reconstructed structures. For the oxygen-covered surface we have
performed a geometry optimization at $1\over 2$, 1, and 2 monolayer
oxygen coverages, and we present results for the equilibrium
configurations, workfunctions and oxygen chemisorption energies. At
half monolayer coverage, we find that oxygen induces a $(1\times 2)$
reconstruction of the surface, while at one monolayer coverage the
chemisorption energy is highest for the unreconstructed surface. Our
results are rationalized by a simple tight-binding description of the
interaction between the O$-2p$ orbitals and the metal valence
states. The resulting bonds are stronger when established with low
coordinated metal atoms, and give rise to an effective
adsorbate-adsorbate interaction when two oxygen atoms are bound to the
same metal orbital. \end{abstract}

\narrowtext


\section{INTRODUCTION}

Oxygen is known to form strong bonds with metal surfaces, and the
oxygen-metal interaction is one of the key elements in many
technologically relevant catalytic reactions, such as the oxidation of
carbon or the reduction of nitrogen from their mono-oxides, occurring
in catalytic mufflers. As a first step towards an understanding of the
microscopic mechanisms responsible for the catalytic activity of some
transition-metal surfaces, we present in this paper a first-principles
study of the interaction between oxygen and the Rh(110) surface. From
the calculations we extract simple chemical models of the metal-oxygen
interaction, which we believe to be of general relevance for
adsorbate-metal systems.

The structure of the oxygen covered Rh(110) surface strongly depends
on the oxygen coverage
\cite{AlRuKiRo93,CoDhCoAsPrRoAtZa93,GiOvErWoScCh93,ScLeWoCh90}.
At very low coverages oxygen is believed
to occupy four-fold long-bridge sites. At a certain critical coverage,
$\theta = \theta_c \lesssim \frac{1}{2}$ monolayers, oxygen starts to
induce a ($1\times 2$) missing-row reconstruction of the surface,
where the oxygen atoms occupy every second three-fold site along the
closed-packed top row. At $\theta={1\over 2}$ the reconstruction is
fully developed, and at higher coverages the surface starts to
deconstruct. At $\theta = 1$ the rhodium substrate is again
unreconstructed and oxygen forms a ($2\times 1$) pattern. Our
calculations, performed at $\theta ={1\over 2}$, 1, and 2 support for
these coverages the picture given above.

From our self-consistent calculations we have extracted a simple
chemical model of the oxygen-metal interaction. From the model we
identify a strong adsorbate-adsorbate interaction when two oxygen
atoms are bound to the same metal orbital, and we discuss this effect in
connection with the calculations at $\theta =1$. We also show that the
model predicts stronger oxygen bonds with low coordinated metal atoms,
and this is the force which drives the oxygen induced reconstruction
at $\theta ={1\over 2}$.

The paper is organized as follows: In Section~\ref{sec:bulk} we
briefly describe the pseudopotential method used for the
calculations, and present our results for bulk Rh; in
Section~\ref{sec:clean} we report on our calculations for the clean
Rh(110) surface, and in Section~\ref{sec:Ocov} on the results for the
oxygen covered surface; in Section~\ref{sec:e1el} we present
the model of the rhodium-oxygen bond; finally, Section~\ref{sec:concl}
contains our conclusions.

\section{Technical details and results for bulk Rh}\label{sec:bulk}

Our calculations are based on density-functional theory within the
local-density approximation (LDA) \cite{HoKo64,KoSh65}. Although the
LDA is known to overestimate cohesive and chemisorption energies, it
has since long proven to be rather reliable as far as relative
energies between different bulk and surface geometries are concerned
\cite{JoGu89}.

The electron-gas data used to implement the LDA are those by Ceperley
and Alder \cite{CeAl80}, as parameterized by Perdew and Zunger
\cite{PeZu81}. The single-particle Kohn-Sham equations are solved
using the plane-wave (PW) pseudopotential method. Isolated surfaces
are modelled by slabs in the supercell geometry. Due to the well known
{\it hardness} of norm-conserving pseudopotentials for first-row
elements---such as O---and, to a lesser extent, for
second-transition-row atoms---such as Rh---, the use of {\it
ultrasoft} (US) potentials \cite{Va90} is mandatory in order to keep
the size of the PW basis sets manageable for the large supercells
necessary to model the surface. The oxygen US pseudopotential is
essentially identical to that of Ref.~\cite{LaPaCaLeVa93} with a core
radius of 1.3 a.u. For rhodium we have used a newly developed variant
of the Vanderbilt scheme \cite{Va90}, where the norm conservation is
released only for those angular-momentum channels which would
otherwise require very hard potentials (the $d$-channel only in the
present case) \cite{St96a}. The pseudopotential is constructed from a
scalar-relativistic all-electron calculation, and it includes the
$4d$-, $5s$- and $5p$-valence states, for which we used the core radii
1.6, 2.53 and 2.53 a.u. respectively. The $4d$ and $5s$ orbitals were
constructed in the atomic reference configuration $4d^85s^1$, while
the $5p$ orbital was taken from the reference configuration
$4d^75s^{0.75}5p^{0.25}$. Basis sets including PW's up to a
kinetic-energy cutoff $E_{cut}\approx 25$--$30$ Ry yield very accurate
results for both elements\cite{LaPaCaLeVa93,St96a}. We finally decided
to adopt a cutoff of 30 Ry. This choice allows one to neglect the
effects of the finiteness of the basis sets even in those instances,
such as {\it e.g.} the calculation of the stress tensor \cite{NiMa85},
which are very sensitive to it. Brillouin-zone (BZ) integrations are
performed using the Gaussian-smearing special-point technique
\cite{MePa89}. We have found that the bulk properties are well
converged using a Gaussian broadening of 0.3 eV and a Monkhorst-Pack
(444) mesh \cite{MoPa76}, which corresponds to 10 $\bf k$ points in
the irreducible wedge of the BZ. In order to test the convergence of
our numerical BZ integrations, in many cases we have made refined test
calculations with a Gaussian broadening of 0.15 eV and a
correspondingly finer mesh. No significant differences were found in
any of the cases examined. This fast convergence with respect to BZ
sampling is due to the relatively smooth Density of States (DOS) at
the Fermi energy, as shown in Fig.~1. In this figure we also show the
projection of the DOS onto individual atomic orbitals. There are 8.0
electrons per atom in the $d$ bands, while the remaining electron has
mixed $s$ and $p$ character.

The calculated lattice constant is $a_0=3.81~\rm\AA$ and the bulk
modulus $B= 3.17~{\rm Mbar}$, in good agreement with the experimental
values (3.80 and 2.76 respectively) \cite{We95}. The overestimation of
the bulk modulus is also found in relativistic all-electron
calculations, and therefore it is not due to the use of
pseudopotentials, but rather due to the LDA\cite{St96a}. The quality
of the agreement between theory and experiment is of the same order as
found for most LDA calculations for metals and semiconductors.

\section{THE CLEAN Rh(110) SURFACE}\label{sec:clean}

The structural properties of the clean unreconstructed and ($1\times
2$) reconstructed surfaces have been studied by minimizing the
supercell total energy with respect to atomic positions, using the
Broyden-Fletcher-Goldfarb-Shanno minimization technique
\cite{Nu92}. The value of the workfunction was estimate by taking the
difference between the calculated Fermi level and the electrostatic
potential in the middle of the vacuum region.

We denote the repeated-cell slab geometry by the notation ($N+M$),
where $N$ is the number of atomic layers in the slab, and $M$ the
thickness of the vacuum region, in units of the interlayer spacing
normal to the surface. Different supercells---going from $(5+5)$ to
($9+5$) and ($7+7$)---where checked in order to test the convergence
of our results. We found no difference between the results obtained
for $M=5$ and $M=7$, and also the dependence upon $N$ is very
small. As for $\bf k$-point sampling we tested a Gaussian broadening
of 0.3 eV and a Monkhorst-Pack (442) mesh \cite{MoPa76}---which
corresponds to 4 $\bf k$ points in the irreducible wedge of the
supercell BZ---and a Gaussian broadening of 0.15 eV in conjunction
with a (882) mesh. We find that the error introduced by using a (442)
mesh is of the same order as the finite size error by using a ($7+5$)
cell.

In Table~\ref{tab:s1x1} we present the calculated atomic positions,
workfunction, surface energy and surface stress for the relaxed
Rh(110) surface, and compare them with the corresponding values
obtained for the unrelaxed surface. The values were obtained from the
($7+5$) cell, with the (442) $\bf k$-point mesh, and the error
estimate was based on calculations performed with the (882) mesh
and/or ($9+5$) supercell geometry. We observe a large inward
relaxation of the first layer, in analogy with what usually occurs at
other open surfaces. However, the calculated relaxation is almost one
time and a half as large as the value obtained from LEED structural
analysis\cite{NiBiHaHeMu87}. Since the calculated bulk modulus is
10\% to large, one could have expected an overestimate of the surface
relaxations, but we also note that the experimental value of
\cite{NiBiHaHeMu87} is probably is too small, since these authors only
included relaxations of the first two surface layers in their
structural model. We find that the surface relaxations strongly
reduce the surface tension, indicating that the surface relaxation and
surface tension are determined by the same driving force. This is in
qualitative agreement with the Effective-Medium Theory of
Ref.~\cite{JaNoPu87,StChJaNo94b}, according to which an inward
relaxation of the outermost atomic plane would decrease the surface
energy by increasing the electron density at surface atoms.

We now turn to the electronic structure of the Rh(110)
surface. In Fig.~\ref{fig:s1x1}a we compare the projections of the DOS
(PDOS) onto the surface atomic layer and onto an atomic layer in the
bulk. The
main difference is the decreased band width and upward shift of the
band center of the first layer PDOS. This effect is in qualitative
agreement with the $\sqrt{N}$ scaling of the band width, where $N$ is
the coordination number, predicted by a simple one-band tight-binding
model. In Fig.~\ref{fig:s1x1}b we show the PDOS of the surface
$d$-orbitals, and we now see the appearance of resonances in the
spectrum caused by the broken metal-metal bonds. The strongest
resonance is caused by the $yz$ orbital, and the significance of this
orbital is that it can only form weak $\delta$ bonds along the
(110) surface row. In Section~\ref{sec:e1el} we will discuss
the chemical bonding between oxygen and the rhodium surface and we will
hint that these resonances give an important contribution to the
chemisorption energy. In analogy to the terminology adopted for
semiconductors, we will call these resonances {\it metallic dangling
bonds}.

Finally, we have considered the $(1\times 2)$ missing-row
reconstructed surface shown in Fig.~\ref{fig:s1x2}. This surface
structure is not found to be stable in nature, but it can be produced in
a meta-stable state by first treating the surface with oxygen which
afterwards is removed with a strong reductant like
CO\cite{CoDhCoAsPrRoAtZa93}. Table~\ref{tab:s1x2} shows the calculated
properties of the relaxed surface, and the reported results were
obtained with a ($7+5$) layer slab and a (882) $\bf k$-point mesh.
Generally the
relaxations are in good agreement with those obtained from the LEED
structural analysis of Ref.~\cite{CoDhCoAsPrRoAtZa93}, and we predict
an increase in the surface energy compared to the unreconstructed
surface of $\approx 0.1~{\rm eV}/(1\times 2~{\rm cell})$, while there
is no significant differences in the surface stress.

\section{The oxygen-covered Rh(110) surface}\label{sec:Ocov}
In this section we present calculations for the Rh(110) surface with a
coverage of $\frac{1}{2}$, $1$, and $2$ oxygen monolayers.
Figure~\ref{fig:all} shows the equilibrium configurations for the
geometries we have considered, and in 
 Table~\ref{tab:all} we report some key values of each
chemisorption geometry. 
  At $\theta = {1\over 2}$ we
find geometry {\it h} to be energetically favoured, while geometry {\it
k} is favoured for $\theta=1$. The structural data predicted by the
present calculations compare well with those determined experimentally
\cite{CoDhCoAsPrRoAtZa93,GiOvErWoScCh93}. The values of the
workfunction and the relative
chemisorption energy of geometries {\it h} and {\it k} also compares
well with the experimental data, however, 
the chemisorption energy is overestimated by $\sim 1.2~\rm eV$, which
 is a typical error of the LDA.

\subsection{Coverage $\frac{1}{2}$: the unreconstructed surface}
\label{ssec:c.5}

We first discuss the stable chemisorption geometries on the
unreconstructed rhodium surface at $\theta = \frac{1}{2}$. For the
calculations we have used a $(2\times1)$ and a $(1\times2)$ surface
cell and for several different initial positions of the oxygen atom,
we have relaxed the structure. We have found six stable
configurations, labeled {\it a, b, c, d, e,} and {\it f} in
Fig.~\ref{fig:all}. The six oxygen chemisorption sites can be divided
into three groups: four-fold-coordinated long-bridge sites (geometries
{\it a} and {\it b}), three-fold-coordinated quasi-long-bridge sites
(geometries {\it c} and {\it d}), and three-fold-coordinated hollow
sites(geometries {\it e} and {\it f}). In the quasi-long-bridge site
the oxygen atom has moved away from the long-bridge position towards a
second-layer atom in order to obtain three-fold coordination.

The chemisorption energy is highest in the three-fold coordinated
hollow site. In these structures most bond lengths between oxygen and
rhodium first-layer atoms are nearly 2 \AA; the bonds with
second-layer atoms are slightly longer for geometries {\it c, d, e},
and {\it f} indicating a weaker bond, while there are no bonds with
second-layer atoms in geometries {\it a} and {\it b}. The main
difference between the structures is in the bond angle between oxygen
and first-layer rhodium atoms, $\theta_{O-1}$. Inspection of
Table~\ref{tab:all} shows that the
chemisorption energy decreases when the bond angle departs from
$90^0$. In section~\ref{sec:e1el} we will provide some arguments which
indicate why this particular angle corresponds indeed to the strongest
oxygen-metal bonds. We also observe that there is almost no inward
relaxation of the surface atoms below the oxygen adsorbate.

Based on a HREELS study of the oxygen vibration frequencies for
different oxygen coverages of the Rh(110) surface, it was proposed in
Ref.~\cite{AlRuKiRo93} that at low coverages oxygen occupies the
long-bridge site. This seems to be in disagreement
with our calculations, since we find the chemisorption energy in
geometry {\it f} to be 0.26 eV higher than in geometry {\it a}. However, our
calculation is for $\theta = \frac{1}{2}$, and at this coverage
adsorbate-adsorbate repulsions might disfavour certain
geometries.

\subsection{Coverage $\frac{1}{2}$: The reconstructed surface}
\label{ssec:O-rec} At $\theta = {1\over 2}$ the annealed surface shows
a $(2\times2)$pg LEED pattern\cite{CoDhCoAsPrRoAtZa93,GiOvErWoScCh93},
indicating that the unit cell is $(2\times2)$ and must have a glide
line in the (110) direction. Furthermore, there is strong
experimental evidence that the underlying rhodium substrate forms a
$(1\times2)$ missing-row structure
\cite{CoDhCoAsPrRoAtZa93,BeCvMoPoToDhLaPrRo93}. If we assume that the
rhodium atom sits in a three-fold site, there are only three
structures consistent with these experimental facts, and these are
geometries {\it h}, {\it i} and {\it j} shown in Fig.~\ref{fig:all}.
The same 
three structures were considered in the LEED IV analysis of
Ref.~\cite{CoDhCoAsPrRoAtZa93}. We find geometry
{\it h} to be energetically favoured, while the oxygen chemisorption
energy is 0.5 and 0.8 eV/atom lower in geometry {\it i} and {\it j},
respectively. Note that the energy ordering of
the three structures indicates that oxygen bonds with low coordinated
rhodium atoms are energetically favoured. In Table~\ref{tab:all} we
compare geometry {\it h} with two independent LEED IV
structural analyses, and the general agreement is rather
good. 

Since oxygen induce the $(2\times1)$ reconstruction on other
transition metal surfaces\cite{BeNo93} we have also considered this
structure. Initially we placed the oxygen atom in an asymmetrical
position above the long bridge site, but after relaxation oxygen
occupied the symmetrical position 0.4 \AA \ above the rhodium surface,
as shown in geometry {\it g} of Fig.~\ref{fig:all}. We now compare the
energetics of this reconstruction with the $(1\times2)$
reconstruction. Table~\ref{tab:all} shows that while each oxygen
atom gains 0.5 eV/atom by forming the $(1\times2)$ reconstruction, the
$(2\times1)$ reconstruction is disfavoured by 0.1 eV/atom. It is
interesting to divide this energy into two contributions, the cost of
reconstructing the substrate ($\Delta \sigma/\Theta$) and the energy
gain by bonding oxygen to the reconstructed surface. The formation of
the $(1\times2)$ reconstruction only costs 0.2 eV/atom (the additional
0.1 eV compared to Table~\ref{tab:s1x2} is due to oxygen induced
strain), while the formation of the $(2\times1)$ reconstruction costs
1.4 eV/atom. By subtracting the formation energy from the
chemisorption energy we see that the oxygen bonding with the
$(2\times1)$ reconstructed surface is 0.6 eV stronger than the bonding
with the $(1\times2)$ reconstructed surface. The stronger bonding is
due to the lower coordination of the surface atoms, and in
section~\ref{sec:e1el} we will identify the electronic origin of this
effect.

Based on the above  calculations we now
discuss a possible  origin of the different reconstructions induced by
nitrogen and oxygen on transition metal (110) surfaces. While oxygen
induces the $(1\times2)$ reconstruction of Rh(110) and the $(2\times1)$
reconstruction of  Cu(110), Ni(110) and Ag(110)\cite{BeNo93}, nitrogen
has almost the 
reverse behaviour, i.e. it induces the $(2\times1)$ reconstruction of
Rh(110)\cite{DhBaCoPrRo95} and a $(1\times3)$ reconstruction of the
Cu(110) and  Ni(110) surfaces where every third (110)
row is missing\cite{NiSpBeCo91,LeDaRo93,VoNiCoCo93}. To understand this behaviour we
have compared the geometry of the theoretical  $(2\times1)$
reconstruction of O/Rh(110), with the experimentally derived
 $(2\times1)$ structure of N/Rh(110)\cite{DhBaCoPrRo95}. The comparison
reveals a rather large difference in the adatom-surface distance; while
oxygen sits 0.4 \AA \ above the surface atoms, there is no height difference 
 between nitrogen and the rhodium surface atoms. We propose that an
adsorbate positioned exactly in between the surface atom in the (100)
row may form the strongest bond, and since nitrogen is a smaller
atom than oxygen, it can better fit in between the rhodium surface atoms.
 This size argument can
also explain the preference of oxygen for the   $(2\times1)$ reconstruction
of silver, since the  lattice constant of silver is 4.09 \AA \
 and the
O-Ag bond length is 2.05 \AA\cite{PuHa84,BeAmScHiPeHa92},
 oxygen fits perfectly in between the silver atoms in the (100) row. A similar picture
is found for copper, where the lattice constant is 3.61 \AA \ and
the O-Cu 
bond length 1.81 \AA\cite{BeNo93}. The lattice constant of nickel  is 3.52 \AA \
and the O-Ni bond length 1.77 \AA\cite{BeNo93}, so in this case the oxygen
atoms do not fit perfectly  in between the nickel atoms, and in
Ref.~\cite{JaNo90} this was identified as a source for lowering the
chemisorption energy. Since nitrogen is a smaller molecule than
oxygen it forms shorter bonds with the metal atoms, and for copper
and nickel the   metal-metal separation gets too large for nitrogen to form
a strong bond with both metal atoms in the (100) row. However, in the
observed  $(2\times3)$ structure on Cu(110) and Ni(110) it is possible
for the surface atoms to relax towards the nitrogen adsorbate, and 
nitrogen can thereby obtain an optimal bond length with the surface atoms.

\subsection{Coverage 1 and 2} \label{ssec:c1}

For $\theta = 1$ we have only studied geometries where the oxygen atom
is three-fold coordinated. We find oxygen to have the highest
chemisorption energy in geometry {\it k} of Fig.~\ref{fig:all}, in
good agreement with experimental findings\cite{GiOvErWoScCh93}.
The oxygen chemisorption site in geometry {\it k} is very similar to
that of geometry {\it f}, and we also find that the chemisorption
energies in the two structures are nearly identical. This suggests
that there is little interaction between the oxygen atoms on the
Rh(110) surface, which is rather surprising compared to the strong
oxygen-oxygen repulsion found on other transition metal surfaces at
similar oxygen coverages\cite{Ki94}. In section~\ref{sec:e1el} we
will show that this different behaviour can be explained by a
substrate mediated adsorbate-adsorbate interaction of electronic
origin, whose strength is related to the filling of the surface bands.

For $\theta = 2$ we have considered two geometries: an oxygen dimer
oriented in the (100) direction
(geometry {\it m} of Fig.~\ref{fig:all}), and an oxygen dimer 
oriented in the (110) direction (geometry {\it n}). We find
geometry {\it m} to be energetically favourable, however, the
chemisorption energy is 1.1 eV lower than in geometry {\it k}. It is
therefore questionable whether this structure can be observed
experimentally after high oxygen exposure, since the structure will
not be energetically favourable compared to geometry {\it k} plus
molecular oxygen.

Geometrically the main difference between geometry {\it k} and
geometry {\it m}, is the shift of the oxygen bond with a second-layer
rhodium atom to an oxygen-oxygen bond. This makes the nature of the
oxygen bonding in geometry {\it m} somewhat intermediate between those
occurring in geometry {\it k} and in molecular oxygen. This structure
might therefore be an important precursor state for oxygen
dissociation.

\section{The chemistry of the rhodium-oxygen bond} \label{sec:e1el}
In this section we will analyse the oxygen chemisorption on rhodium in
terms of a simple tight-binding description. A tight-binding model of
chemisorption has been put forward by several authors\cite{Ne69,Ho88,No89b}.

The first step in building a tight-binding description of oxygen
chemisorption is to identify the metal and oxygen orbitals which are
responsible for the bond formation. For the present problem these
consist of all the valence orbitals of rhodium (4$d$, 5$s$ and 5$p$)
and the O-${2p}$ orbitals, while we can disregard the O-${2s}$
orbital, since its resonance is $\approx$ 19 eV below the rhodium
Fermi level. As in the model of Refs.~\cite{JaNo90,HaNo95} we describe
the interaction between the O-${2p}$ orbitals and the metal orbitals
in two steps: First the unperturbed orbitals interact with the rhodium
Rh-${5s}$ and Rh-${5p}$ orbitals, and then the resulting
renormalized O-${2p}$ orbitals interact with Rh-${4d}$ orbitals.  The
different steps are illustrated in Fig~\ref{fig:ochem}. For the
unperturbed O-${2p}$ state we take the atomic eigenvalue and use the
Rh(110) workfunction to position it relative to the rhodium Fermi
level ($\epsilon_{2p}-\epsilon_f=-4.1$ eV). We note that the
corresponding unperturbed O-${2s}$ eigenstate is positioned $-18.6$ eV
relative to the Fermi level, while the fully selfconsistent
calculation of geometry {\it f} has a O-${2s}$ resonance at $-18.9$
eV. The interaction between oxygen and the broad metal $sp$ bands is
similar to the interaction between oxygen and a Jellium surface. This
interaction is well described by the weak coupling limit of the
Newns-Anderson model, and gives rise to a broadening and a shift of
the O-${2p}$ level. The interaction between the renormalized O-${2p}$
orbitals and the narrow $d$ band is, on the other hand, described by
the strong coupling limit of the Newns-Anderson model, and gives rise
to a splitting of the levels into bonding--anti-bonding states,
similar to the interaction between two atomic orbitals.

By dividing the oxygen-metal interaction into an $sp$ and a $d$ part
we obtain a simple picture of the oxygen chemisorption energy on
different transition metal surfaces. The interaction with the metal
$sp$ bands gives a large contribution to the chemisorption energy,
however, due to the similarity of the transition metal $sp$ bands this
contribution is more or less constant throughout the transition
series. The $d$ band contribution, on the other hand, decreases with
the filling of the band, and for the noble metals the O-${2p}$ -- $d$
anti-bonding level will be filled\cite{HaNo95} and in this case there
will be a O-${2p}$ -- $d$ repulsion due to
orthogonalization. Figure~\ref{fig:ochem}c schematically shows that
the O-${2p}$ -- Rh-${4d}$ anti-bonding level is well above the Fermi
level, and the interaction with the rhodium $d$-band, therefore, gives
a large contribution to the chemisorption energy. We note that a
substantial part of the O-${2p}$ --
Rh-${4d}$ bond is formed between the O-${2p}$ orbitals
and localized Rh-${4d}$ states with energies close to the Fermi level,
i.e. metallic dangling bonds. This is illustrated in
Figure~\ref{fig:pdorb} where we
show the Rh-${4d_{yz}}$ PDOS for both the clean and oxygen covered Rh(110)
surface. The figure shows that the ${4d_{yz}}$ dangling bond
resonance at $\epsilon_d$ interacts with the renormalized adsorbate
adsorbate level at $\epsilon_a$ and forms a bonding state at
$\epsilon_a-\Delta_1$ and an anti-bonding state at
$\epsilon_d+\Delta_1$, where $\Delta_1=
\protect\sqrt{\Delta_0^2+V^2}-\Delta_0$, $\Delta_0 =
(\epsilon_d-\epsilon_a)/2$, and $V$ is the O-$2p$ -- Rh-$4d$ coupling
matrix element. 

Since the O-${2p}$ level is below the Rh-${4d}$ band centre, the
bonding state will have mostly O-${2p}$ character. There will
therefore be a charge transfer from the Rh-${4d}$ orbitals into the
O-${2p}$ orbitals. This charge transfer can be seen in
Fig.~\ref{fig:chargediff}, which shows the charge-density difference
between the oxygen+rhodium system and the two separated systems. The
plot shows a charge transfer from rhodium orbitals of symmetry
$d_{3z^2-r^2}$(with $z$ along the bond axes) to the O-${2p}$
orbitals. We note that the rhodium $d_{3z^2-r^2}$ orbitals have the
largest overlap with the the O-${2p}$ orbitals, and therefore give
rise to the strongest rhodium-oxygen bonds.

In Fig.~\ref{fig:oint} we show the O-${2p}$ PDOS of geometries {\it f}
and {\it k}, and in the spectrum of geometry {\it k} we can identify 3
resonances. Apparently this is in conflict with the model developed in
the previous section which only predicted two resonances,
corresponding to the bonding and anti-bonding state of two interacting
atomic orbitals. However, in geometry {\it k} two O-${2p}$ orbitals
couple to the same Rh-${4d}$ orbital, and the bonding is therefore
related to the interaction between three atomic orbitals, which give
rise to a bonding resonance at $\epsilon_a-\Delta_2$, an anti-bonding
resonance at $\epsilon_d+\Delta_2$, and a non-bonding resonance at
$\epsilon_a$, where $\Delta_2=
\protect\sqrt{\Delta_0^2+2 V^2}-\Delta_0$. From the
positions of the three resonances($-6.8$ eV, $-5.2$ eV, and $0.8$ eV)
we can determine the parameters of the atomic model. We find
the renormalized adsorbate level $\epsilon_a=-5.2$ eV, the Rh-${4d}$ level
$\epsilon_d=-0.8$, and the O-${2p}$--Rh-${4d}$ coupling matrix element
$V=2.2$ eV. Compared to the unperturbed adsorbate level
($\epsilon_{2p}-\epsilon_f=-4.1$ eV), the rhodium 4$d_{yz}$ resonance
of Fig.~\ref{fig:pdorb}, and the O-${2p}$ -- Rh-${4d}$ coupling matrix
element obtained from Harrisons solid-state table\cite{Ha89} $V_{\rm
Har}=2.0$ eV, these values seem indeed reasonable. We also note that
for two interacting atomic orbitals the parameters predict a bonding
resonance at $-6.1$ eV, in good agreement with the O-${2p}$ PDOS of
geometry {\it f}.

The above chemisorption model gives rise to a strong
adsorbate-adsorbate interaction, which can be either attractive or
repulsive depending on the position of the metal state $\epsilon_d$
relative to the Fermi level. If the metal state is empty
($\epsilon_d>\epsilon_f$) the $d$-band contribution to the
chemisorption energy of non-interacting adsorbates is $E_1=2\Delta_1$,
while for interacting adsorbates it is $E_2=\Delta_2$, and since $\Delta_1>\Delta_2/2$ there is a
repulsive adsorbate-adsorbate interaction. However, if
$\epsilon_f>\epsilon_d+\Delta_1$ then $E_1=0$ and
$E_2=\Delta_2-\epsilon_f$, and in this case there is an
adsorbate-adsorbate attraction. The crossover between repulsive and
attractive interaction happens in the range
$\epsilon_d<\epsilon_f<\epsilon_d+\Delta_1$, where the difference in
the $d$-band contribution to the chemisorption energy is: $E_{2}-E_{1}
=( \epsilon_d+\Delta_2 -\epsilon_f) - 2 (\epsilon_d+\Delta_1
-\epsilon_f)$. From the model parameters we estimate an interaction
energy of $E_{2}-E_{1}=0.6$ eV, and since two orbitals per oxygen atom
contribute to the interaction, the $d$-band contribution to the
chemisorption energy is 1.2 eV larger in geometry {\it k} than in
geometry {\it f}. However, the model also predicts an
adsorbate-adsorbate repulsion caused by the O-${2p}$ interaction with
the almost empty $sp$-band, and since we know from the full
selfconsistent calculation that the oxygen chemisorption energy is the
same in the two structures, this $sp$-mediated repulsion must
compensate the $d$-mediated attraction. In geometry {\it l} of
Fig.~\ref{fig:all} two neighboring oxygen atoms can only couple to the
same metal $s$ orbital, and there is only an $s$-mediated repulsion,
which gives rise to a 0.2 eV lower chemisorption energy than in
geometry {\it l}.

Work is in progress to make the model more quantitative, and
preliminary results show that the model can account for the different
adsorbate-adsorbate interaction on transition metal and aluminum
surfaces\cite{St96b}.

In Section~\ref{sec:clean} we showed that the bandwidth of the atom
PDOS scale as $\sqrt{N}$, where $N$ is the coordination of the atom,
and that there is a corresponding shift of the band centre which
conserves the filling of the band. This was illustrated for the
Rh(110) surface in Fig.~\ref{fig:s1x1}, and in the present context we
note that the rhodium band centre is shifted upwards. This shift gives
rise to a similar shift of the rhodium--oxygen anti-bonding state, and
since the bond energy is given by the distance between the
anti-bonding state and the rhodium Fermi level the bond is
strengthened. The energy gain by the oxygen induced $(1\times2)$
reconstruction is therefore due to the oxygen bond with the the second
layer rhodium atom, which on the reconstructed surface has
coordination 9 and on the unreconstructed surface coordination 11.

\section{Conclusions}\label{sec:concl}
In this paper we presented detailed first-principles calculations for
the clean and half, one and two monolayer oxygen-covered Rh(110)
surface. At half monolayer
oxygen coverage
we find a $(1\times2)$ reconstruction of the surface,
contrary to the oxygen induced $(2\times1)$ reconstruction of Cu(110),
Ag(110) and Ni(110), but in good agreement with experimental studies
of O/Rh(110). Calculations for the theoretical $(2\times1)$ O/Rh(110)
structure suggest that this difference in behaviour is related to the
scale between the lattice constant and the oxygen-metal
bond. Generally the calculations show that the oxygen chemisorption
can be rationalized in terms of a tight-binding model which include
the O-${2p}$ interaction with the rhodium valence states. This model
give rise to a short ranged but strong adsorbate-adsorbate interaction
which can be either repulsive or attractive dependent of the filling
of the surface bands.

\acknowledgements We gratefully acknowledge many discussions with 
K.~W. Jacobsen, J.~K. N\o rskov, R. Rosei, and E. Tosatti. Thanks are
due to R. Valente for porting our plane-wave ultra-soft
pseudopotential program to an IBM-SP2 parallel machine, and
D. Vanderbilt for providing an atomic program for the generation of
ultra-soft pseudopotentials. Kurt Stokbro acknowledges financial
support from the European Union through HCM contracts ERBCHBGCT 920180
and ERBCHRXCT 930342. This work has been partially supported by the
Italian {\it Consiglio Nazionale delle Ricerche} within the {\it
Supaltemp} project.

\appendix
\section{Calculation of the chemisorption energy}
\label{app:chem}
To estimate the chemisorption energy relative to the O$_2$ molecule we
have used essentially the same procedure as
Ref.~\cite{JaHaJaNo95}. With the non-spinpolarized pseudopotential
code we have calculated the energy of the O/Rh(110) system
($E_{O,Rh}$) and the surface energy ($E_{Rh}$) using the same
super cell and $k$-point sampling. To find the reference molecular energy we
have used an atomic program to calculate the 
energy $E_{O}^{pol}$ of a spin-polarized oxygen atom, and from that
subtracted the O$_2$ atomization energy $E_b=3.74$ eV/atom found by Becke\cite{Be92}.
This value has to be corrected for the use of a finite basis set
in the plane-wave calculation, which we estimate to be
$E_{pw}=0.05$ eV/atom. From these energies we calculate the
chemisorption energy, $E_{chem}$, by
\begin{equation}
E_{chem} = E_{O}^{pol}-E_b+E_{pw} +E_{Rh} - E_{O,Rh}.
\end{equation}

\begin{figure}\caption{The rhodium bulk DOS (solid line) and its
projection onto $sp$ and $d$ atomic orbitals (dotted
lines). \label{fig:dos}} \end{figure}

\begin{figure} \caption{The DOS of the Rh(110) surface projected onto: a)
layer 1 (solid line) and layer 4 (dashed line). b) The five $d$ orbitals
of layer 1. \label{fig:s1x1}} \end{figure}

\begin{figure} \caption{The geometry of the $(1\times2)$ missing row
reconstruction. \label{fig:s1x2}} \end{figure}

\begin{figure} \caption{Oxygen binding geometries on the Rh(110)
surface at coverage $\frac{1}{2}$, 1 and 2. The energetically
favourable configurations at the three coverages are geometry {\it h},
{\it k} and {\it m},\label{fig:all}} \end{figure}

\begin{figure} \caption{a) The atomic O-${2p}$ eigenstate. b)
Schematic O-${2p}$ PDOS for an oxygen atom coupling to the rhodium
$5s$ and $5p$ orbital. c) the O-${2p}$ PDOS of geometry~f. d) the
first layer PDOS of the clean Rh(110) surface.\label{fig:ochem}}
\end{figure}

\begin{figure}\caption{The DOS of geometry {\it f} projected onto the the
O-${2p}$ orbital and rhodium $4d_{yz}$ orbital. The solid lines show
the PDOS of the interacting oxygen-rhodium system, while the dashed
lines show the PDOS of isolated oxygen and rhodium
systems.\label{fig:pdorb}}
 \end{figure}

\begin{figure} \caption{Contour plot of the oxygen induced
charge-density difference in geometry {\it f}. Dashed contour lines indicate
charge depletion. The top view shows the charge density difference in
the plane defined by the two first layer rhodium atoms(large grey
circles) and the oxygen atom(small solid circle). The side view shows
the plane defined by the first and second layer rhodium atoms and the
oxygen atom. \label{fig:chargediff}} \end{figure}

\begin{figure} \caption{The O-${2p}$ PDOS of geometry {\it f} and {\it
k}. The resonances in geometry
{\it f} can be described by the bond formation between two levels,
$\epsilon_a$ and $\epsilon_d$, coupled by $V$. In geometry {\it k} the
resonances can be described by the bond formation between three
levels, where two levels at $\epsilon_a$ are coupled by $V$ to a level
at $\epsilon_d$. The level shifts are given by $\Delta_1=
\protect\sqrt{\Delta_0^2+V^2}-\Delta_0$ and $\Delta_2=
\protect\sqrt{\Delta_0^2+2V^2}-\Delta_0$, where $\Delta_0 =
(\epsilon_d-\epsilon_a)/2$.\label{fig:oint}} \end{figure}

\widetext

\begin{table}
\caption{Interlayer relaxations ($\Delta d_{12}$, $\Delta
d_{23}$, $\Delta d_{34}$), work function (W), surface energy
($\sigma$), 
and surface tensions ($\sigma_{xx}$,$\sigma_{yy}$) of the Rh(110)
surface (the $x$ axis is oriented in the (110) direction).
The first row shows the available experimental data, while the
second row shows the calculated values before surface relaxation, and
the the third row after relaxation. The numbers in parenthesis indicate an
estimate of the numerical error on the last displayed digit,
due to the super-cell size and
$k$-point sampling.\label{tab:s1x1}}
\begin{tabular}{lllclccc}
&$ \Delta d_{12}$(\%) & $\Delta d_{23}$(\%) & $\Delta d_{34}$(\%) &
W(eV) & $\sigma$(eV/\AA$^2$) 
& $\sigma_{xx}$(eV/\AA$^2$) & $\sigma_{yy}$(eV/\AA$^2$) \\
\tableline
Expt. & $-7(1)\tablenotemark[1]$ & $2(1)\tablenotemark[1]$ & &
$5.1\tablenotemark[2] $ \\
LDA unrelaxed & & & & 5.13(5) & 0.198(5) & 0.28(2) & 0.25(2) \\
LDA relaxed & $-9.7(5)$ & 3.6(8) & $-2.2(9)$ & 5.14(5) & 0.191(5) & 0.20(2) &
0.12(2) \\
\end{tabular}
\tablenotetext[1]{From Ref. \cite{NiBiHaHeMu87}}
\tablenotetext[2]{From Ref. \cite{Co91}}
\end{table}

\begin{table}
\caption{Experimental and calculated surface properties of the
$(1\times2)$ reconstructed 
surface. For the definition of the symbols see
Table~\protect\ref{tab:s1x1} and Fig.~\protect\ref{fig:s1x2}. The
workfunction, surface energy and surface tension are given relative to
the unreconstructed relaxed surface.\label{tab:s1x2}}
\begin{tabular}{llllllcccc}
&$\delta_y$ & $\Delta d_{12}$ & $\Delta d_{23}$ &
$\Delta d_{34}$ &$\Delta d_{45}$ & $\Delta W$ & $\Delta \sigma$
& $\Delta \sigma_{xx}$ & $\Delta \sigma_{yy}$\\
& (\%) & (\%) & (\%) & (\%) & (\%) & (eV) & (eV/\AA$^2$) &
(eV/\AA$^2$) & (eV/\AA$^2$) \\
\tableline
Expt. & 0.0\tablenotemark[1] &$-11.4\tablenotemark[1]$ &
$-7.7\tablenotemark[1]$ & 11.2$^a$ & $-5.5\tablenotemark[1]$ & \\ 
LDA & 0.6(2) & $\ -7.1(5)$ & $-6.7(5)$ & \ 6(1) & $-3(1)$ & 0.16(3) &
0.004(1) & 0.00(2) & 0.01(2)\\
\end{tabular}
\tablenotetext[1]{From Ref. \cite{CoDhCoAsPrRoAtZa93}}
\end{table}

\begin{table}
\caption{Some key quantities describing the oxygen bonding in the
geometries of Fig.~\protect\ref{fig:all}. The reported values are: The
chemisorption energy $E_{chem}$, substrate surface energy $\Delta 
\sigma/\theta$, workfunction ($\Delta W$), oxygen dipole ($\mu$),
oxygen-surface distance ($z_{O-Rh_1}$), the shortest bond length with
a first layer atom ($d_{O-1}$), the shortest bond length with a second
layer atom ($d_{O-2}$), the bond angle between bonds with first layer
atoms ($\theta_{O-1}$), the bond angle between bonds with a
first layer atom and a second layer atom ($\theta_{O-2}$), and the
first layer relaxation of the rhodium substrate($\Delta d_{12}$) (for
the geometries {\it a, b, c, d, e,} and {\it f} the surface atoms
relax differently and we show the value of both the smallest and the
largest relaxation). Note that in geometry~{\it i} the angle $\theta_{O-1}$ is
between the oxygen bonds with second layer atoms, and in
geometry~{\it j} the bond lengths and bond angles are reported for the
oxygen bonds with second and third layer atoms. The  oxygen dipole-moment was derived from the
oxygen induced workfunction change, and
the calculation of the chemisorption energy is described in Appendix~\ref{app:chem}.\label{tab:all} } 
\begin{tabular}{l|llll|lllllll}
 &\multicolumn{1}{c}{$ E_{chem}$} & $\Delta \sigma/\Theta$ & $\Delta W$ &
 \multicolumn{1}{c}{$\mu$} & $z_{O-1}$ & $d_{O-O}$ & $d_{O-1}$ & $d_{O-2}$ & 
$\theta_{O-1}$ & $\theta_{O-2}$ &\multicolumn{1}{c}{ $\Delta d_{12}$}
\\
& (eV/O)& (eV/O) & (eV) & (e~\AA/O) & \multicolumn{1}{c}{(\AA)} &
\multicolumn{1}{c}{(\AA)} & \multicolumn{1}{c}{(\AA)} & \multicolumn{1}{c}{(\AA)} & & &\multicolumn{1}{c}{ (\%)} \\
\tableline
{\it a)} & 1.97(4) & 0.23 & 0.9(1) & 0.11(1)& 0.94(4) & 2.69(1) & 1.96(2) & 2.77(3) & 123(2)$^0$ &
65(2)$^0$ & $-15/10 $\\ 
{\it b)} & 1.86(4) & 0.15 & 0.50(5) & 0.056(5) & 0.57(3) & 3.81(1) & 1.99(2) & 2.36(3) & 147(3)$^0$ &
77(3)$^0$ & $-13/3 $\\ 
{\it c)} & 1.89(4) & 0.05 & 0.51 (5) &0.057(5) & 0.62(3) & 2.69(1) & 2.06(2) & 2.07(2) & 131(2)$^0$ &
80(2)$^0$ & $-14/\! - \! 2$ \\
{\it d)} & 1.94(4) & 0.16 & 0.38(2) & 0.044(3) & 0.53(1) & 3.81(1) & 2.03(1) & 2.09(1) & 140(1)$^0$ &
82(1)$^0$ & $-17/0 $\\ 
{\it e)} & 2.03(3) & 0.19 & 0.40(4) & 0.046(4)& 0.59(2) & 2.69(1) & 2.01(1) & 2.02(1) & 103(2)$^0$ &
87$^0$ & $-11/ 2 $\\ 
{\it f)} & 2.23(2) & 0.11 & 0.44(5) & 0.050(5)& 0.71(1) & 3.81(1) &
2.00(1) & 2.05(1) & \ 93(2)$^0$ &
82(1)$^0$ & $-10/\! - \! 3$\\ 
{\it g)} & 2.18 & 1.42 & 0.50 & 0.06 & 0.50 & 3.81 & 1.96 & 2.26 &
151$^0$ & 79$^0$ & \ $ -1 $\\
{\it h)} & 2.75 & 0.19 & 0.85 & 0.10& 0.60 & 3.78 & 2.00 & 2.00 & \
85$^0$ & 85$^0$ & \ $-3 $ \\ 
Expt. & 1.54\tablenotemark[1] & &
0.65\tablenotemark[2] & 0.074\tablenotemark[2]& 0.54\tablenotemark[3] & 3.82\tablenotemark[3] & 1.98\tablenotemark[3] & 1.97\tablenotemark[3]
& \ 86$^0$\tablenotemark[3] & 88$^0$\tablenotemark[3] & \ $-1 $\tablenotemark[3]\\ 
 Expt. & & & & & 0.50\tablenotemark[4] & 3.89\tablenotemark[4] & 2.01\tablenotemark[4] &
1.88\tablenotemark[4] & \ 84$^0$\tablenotemark[4] &
87$^0$\tablenotemark[4] &\ $-3 $\tablenotemark[4]\\ 
{\it i)} & 2.26 & 0.19 & 0.84 & 0.10 & 0.31 & 4.44 & 2.02 & 2.02 & (85$^0$) & 82$^0$ & $-12$ \\ 
{\it j)} & 2.00 & 0.19 & 0.14 & 0.016& $\!\!\!\!\!-0.61$ & 2.98 & $\!\!(2.02)$ & $\!\!(2.06)$ & (87$^0$) & $\!\!(84^0)$ & \ $-6$ \\
\tableline
{\it k)} & 2.23(2) & 0.06 & 0.87(4) & 0.049(3) & 0.66(1) & 2.96(1) &
1.99(2) & 2.06(2) & \ 85(1)$^0$ &
 82(1)$^0$ & \ $-3 $\\ 
 Expt. & 1.06\tablenotemark[1] & & 0.73\tablenotemark[2] &
0.041\tablenotemark[2] & 0.60\tablenotemark[3] & 2.93\tablenotemark[3] & 1.98\tablenotemark[3] & 2.05\tablenotemark[3]
& \ 86$^0$\tablenotemark[3] & 84$^0$\tablenotemark[3] & \ \ \ 1\tablenotemark[3] \\ 
{\it l)} & 2.06(2) & 0.10 & 0.6(1) & 0.034(5)& 0.60(2) & 2.69(1) &
2.00(3) & 2.03(3) & \ 85(2)$^0$ &
85(2)$^0$ &\ \ \ 0 \\ 
\tableline
{\it m)} & 1.1(1) & 0.09 & 1.4(1) & 0.040(4)& 1.32(1) & 1.32(1) &
2.26(1) & 2.84(2) & \ 73(1)$^0$ & 64(1)$^0$ &\ \ \ 7\\ 
{\it n)} & $\!\!\!\!\!-1.1$(1) & 0.03 & $\!\!\!\!\!-0.65$ & & 1.81 & 1.15 & 2.74 & 3.16 & \ 88$^0$ & 53$^0$ & \ $-4$ \\
\end{tabular}
\tablenotetext[1]{Ref. \cite{CoDhKiPaPaPrRo92}}
\tablenotetext[2]{Ref. \cite{ScLeWoCh90}}
\tablenotetext[3]{Ref. \cite{GiOvErWoScCh93}}
\tablenotetext[4]{Ref. \protect\cite{CoDhCoAsPrRoAtZa93} }
\end{table}

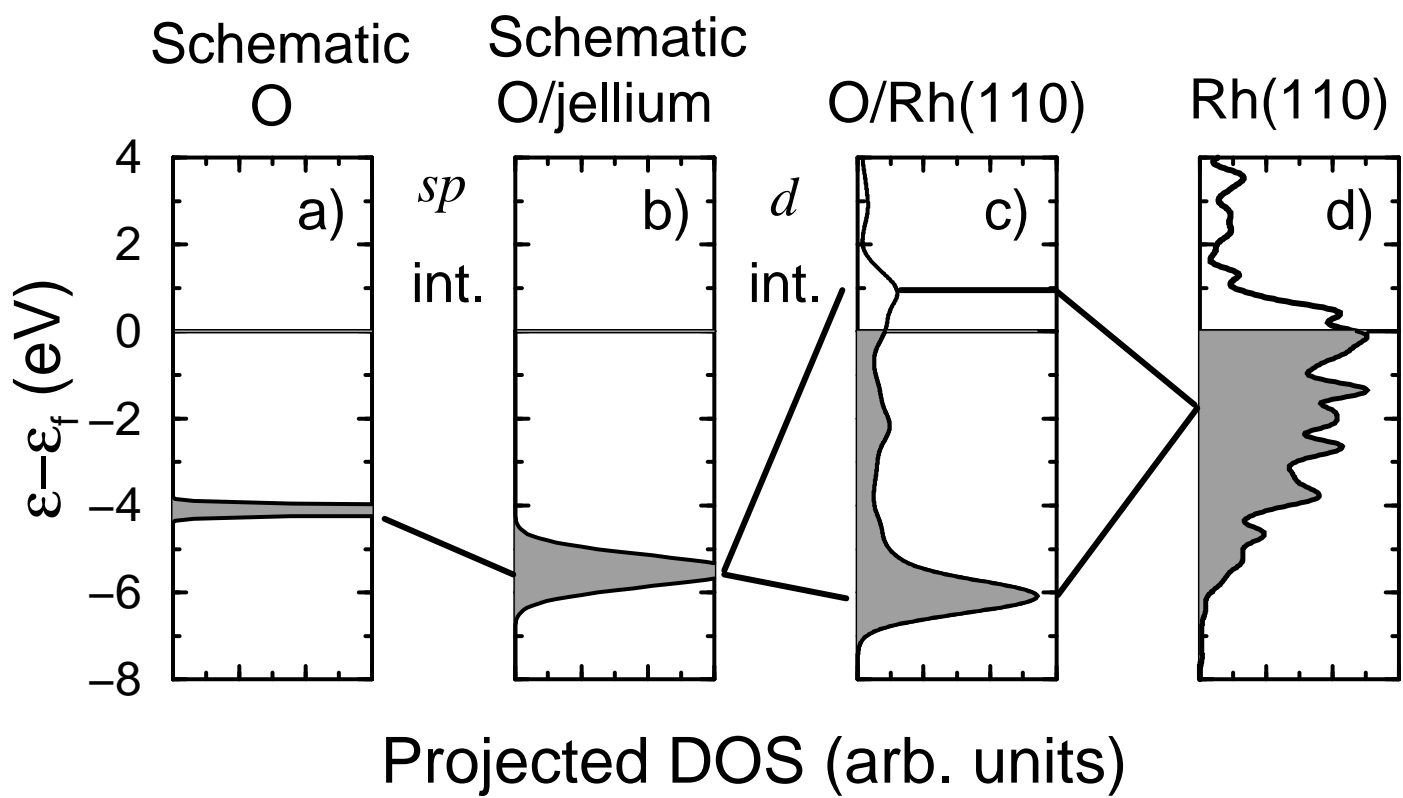
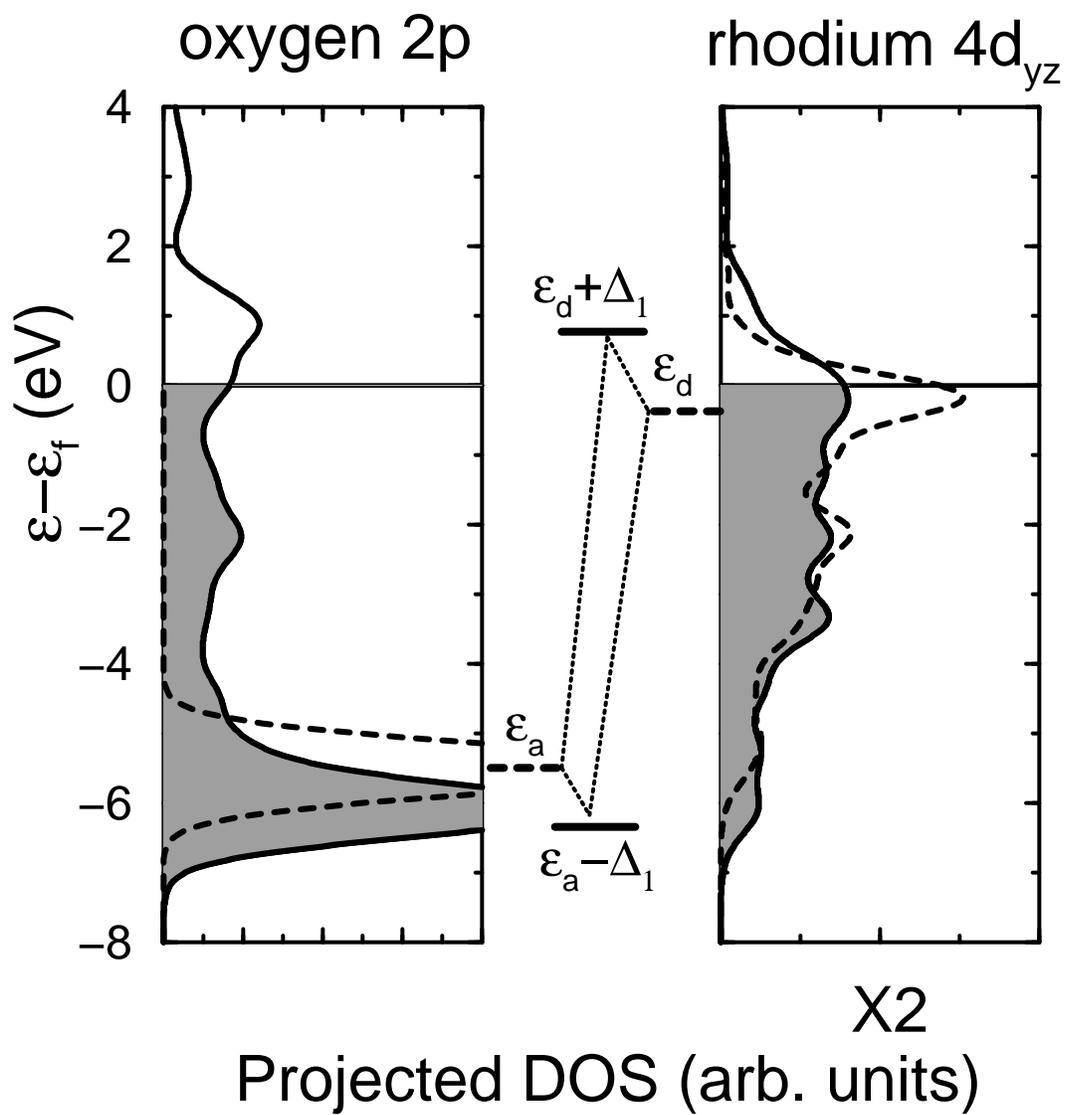
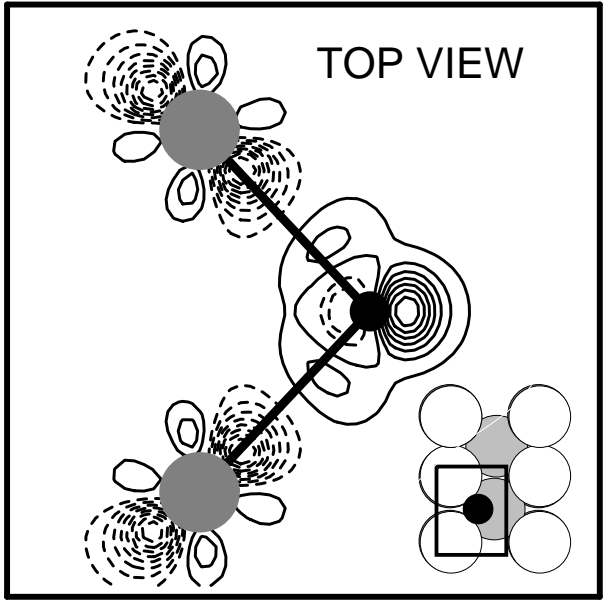
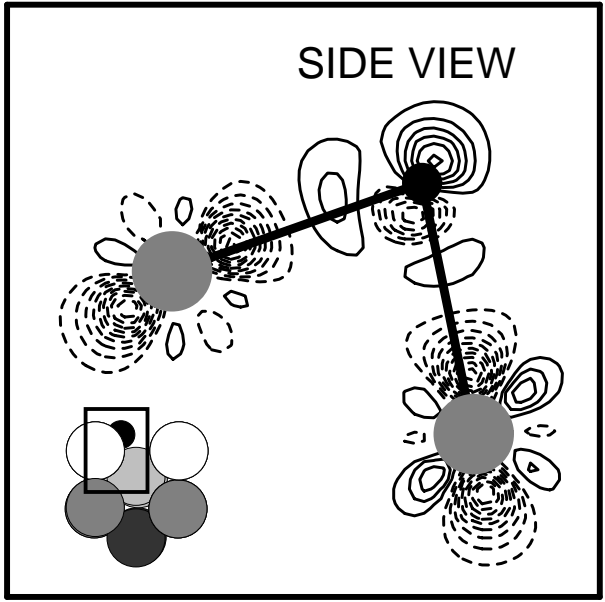
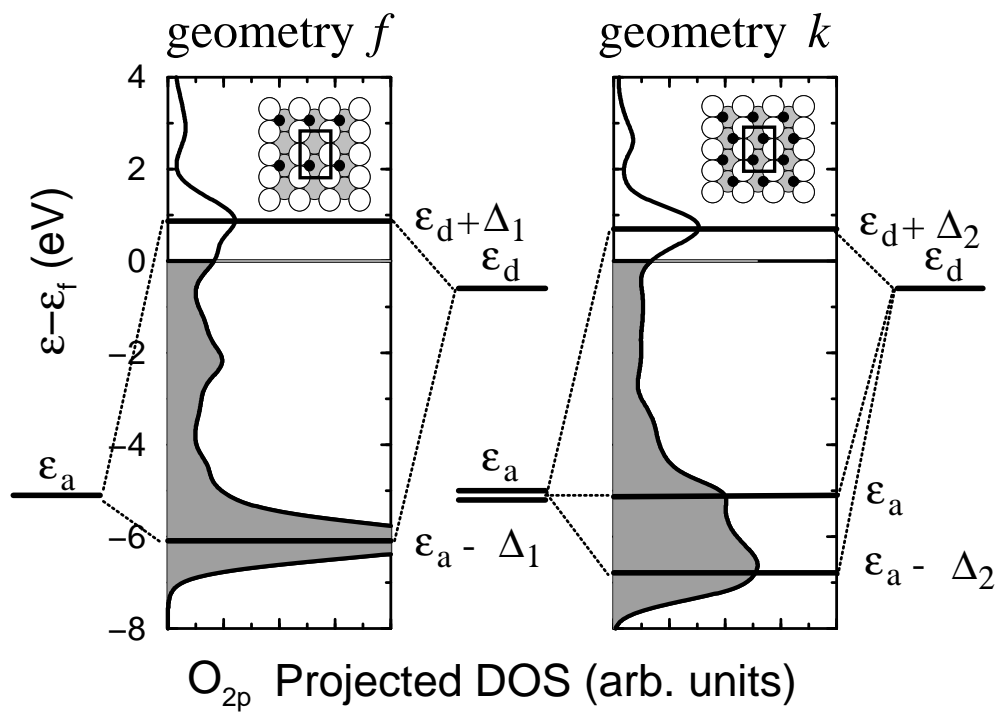
\end{document}